\documentclass[12pt]{article}
\usepackage[english,german,french,polish]{babel}
\usepackage[T1]{fontenc}

\begin{document}

\hfill IFT-01/14

\baselineskip 0.75cm
\topmargin -0.6in
\oddsidemargin -0.1in

\let\ni=\noindent

\renewcommand{\thefootnote}{\fnsymbol{footnote}}

\newcommand{\CKM}{Cabibbo--Kobayashi--Maskawa }

\pagestyle {plain}

\setcounter{page}{1}

\pagestyle{empty}

~~~

\begin{flushright}

\end{flushright}

\vspace{0.3cm}

{\large\centerline{\bf LSND effect as a Chooz-restricted "\,$\!$sterile" perturbation }}

\vspace{0.2cm}

{\large\centerline{\bf of three-neutrino texture{\footnote {Supported in part by the Polish State Committee for Scientific Research (KBN), grant 5 PO3B 119 20 (2001-2).}}}}

\vspace{0.3cm}

{\centerline {\sc Wojciech Kr\'{o}likowski}}

\vspace{0.3cm}

{\centerline {\it Institute of Theoretical Physics, Warsaw University}}

{\centerline {\it Ho\.{z}a 69,~~PL--00--681 Warszawa, ~Poland}}

\vspace{1.2cm}

{\centerline{\bf Abstract}}

\vspace{0.3cm}

Considering the hypothesis of mixing of three active neutrinos with, at least, one sterile neutrino, we report on a simple $4\times 4$ texture whose $ 3 \times 3 $ part arises from the popular bimaximal texture for three active neutrinos  $ \nu_e\,, \,\nu_\mu\,, \,\nu_\tau $, where $ c_{12} = 1/\sqrt{2} = s_{12}$, $ c_{23} = 1/\sqrt{2} = s_{23}$ and $ s_{13} = 0$. Such a $3\times 3 $ bimaximal texture is perturbed through a rotation in the 14 plane, where $\nu_4 $ is the extra neutrino mass state induced by the sterile neutrino $\nu_s $ which becomes responsible for the LSND effect. Then, with $m^2_1\simeq m^2_2$ we predict that $\sin^2 2\theta_{\rm atm} = \frac{1}{2}(1+ c^2_{14}) \sim 0.99$ and $\sin^2 2\theta_{\rm LSND} = \frac{1}{2}s^4_{14} \sim 4.5\times10^{-4}$, and in addition $ \Delta m^2_{\rm atm} = \Delta m^2_{32}$ and $ \Delta m^2_{\rm LSND} = |\Delta m^2_{41}|$,  where $c^2_{14} = \sin^2 2\theta_{\rm sol} \sim 0.97$ and $\Delta m^2_{21} = \Delta m^2_{\rm sol} \sim 10^{-7}\;{\rm eV}^2$ if {\it e.g.} the LOW solar solution is applied. In this four-neutrino texture with $m^2_1 \simeq m^2_2 $ the sum rule $\sin^2 2 \theta_{\rm sol} + \frac{1}{2}\sin^2 2 \theta_{\rm Chooz} + \sin^2 2 \theta_{\rm LSND}  = 1$ holds in the two-flavor approximation (for each of three cases), leaving room for the LSND effect, depending on the magnitude of Chooz effect that, not observed so far, leads (at present) to the upper bound $\sin^2 2\theta_{\rm LSND}\stackrel{<}{\sim} 1.3\times10^{-3}$ and the lower bound $\sin^2 2\theta_{\rm sol} \stackrel{>}{\sim} 0.95$. At the end a four-neutrino seesaw mechanism is sketched.

\vspace{0.6cm}

\ni PACS numbers: 12.15.Ff , 14.60.Pq , 12.15.Hh .

\vspace{0.8cm}

\ni May 2001

\vfill\eject

~~~
\pagestyle {plain}

\setcounter{page}{1}

\vspace{0.2cm}

The present status of experimental data for atmospheric $\nu_\mu $'s as well as solar $ \nu_e $'s favours oscillations between three conventional neutrinos $ \nu_e\,, \,\nu_\mu\,, \,\nu_\tau $ only [1]. However, the problem of the third neutrino mass-square difference, related to the possible LSND effect for accelerator $\nu_\mu $'s, is still actual [2], stimulating a further discussion about mixing of these three active neutrinos  with, at least, one hypothetical sterile neutrino $ \nu_s $ (although such a sterile neutrino is not necessarily able to explain the LSND effect [3]). As a contribution to this discussion, we report in this note on a simple $4\times 4$ texture for three active and  one sterile neutrinos, $ \nu_e\,, \,\nu_\mu\,, \,\nu_\tau $ and $ \nu_s $, whose $3\times 3$ part arises from the popular bimaximal texture [4] working {\it grosso modo} in a satisfactory way for solar $\nu_e $'s and atmospheric $\nu_\mu $'s if the LSND effect is ignored. Such a $3\times 3$  bimaximal texture is perturbed [5] by the sterile neutrino $\nu_s$ inducing one extra neutrino mass state $\nu_4$ and so, becoming responsible for the possible LSND effect. In fact, with the use of our $4\times 4$ texture we predict that $\sin^2 2\theta_{\rm LSND} = \frac{1}{2} s^4_{14}$ and $ \Delta m^2_{\rm LSND} = |\Delta m^2_{41}|$, while $\sin^2 2\theta_{\rm sol} = c^2_{14}$ and $\Delta m^2_{\rm sol} = \Delta m^2_{21}$ as well as  $\sin^2 2\theta_{\rm atm} = \frac{1}{2}(1+ c^2_{14})$ and $ \Delta m^2_{\rm atm} = \Delta m^2_{32}$, if $m^2_1 \simeq m^2_2$ (and both are different enough from $m^2_3$ and $m^2_4$). Here, $c^2_{14} \sim 0.97$ and $\Delta m^2_{21} \sim 10^{-7} \,{\rm eV}^2$ if {\it e.g.} the LOW solar solution [1,6,7] in its recent estimation [7] is accepted; then we predict $\sin^2 2 \theta_{\rm atm} \sim 0.99$ and $\sin^2 2\theta_{\rm LSND} \sim 4.5\times10^{-4} $. 
 
In the popular $3\times 3$ bimaximal texture the mixing matrix has the form [4]

%Eq. 1
\begin{equation} 
U^{(3)} =   \left( \begin{array}{ccc} 1/\sqrt{2} & 1/\sqrt{2} & 0 \\
- 1/{2}\, & \;\,1/2 & 1/\sqrt{2} \\ \;\,1/2 & -1/2\, & 1/\sqrt{2}\end{array} \right) \;.
\end{equation} 

\ni Such a form corresponds to $ c_{12} = 1/\sqrt{2} = s_{12}$, $ c_{23} = 1/\sqrt{2} = s_{23}$ and $ s_{13} = 0$ in the notation used for a generic \CKM$\!\!$--type matrix [8] (if the LSND effect is ignored, the upper bound $|s_{13}| \stackrel{<}{\sim} 0.1 $ follows from the negative result of Chooz reactor experiment [9]). Going out from the form (1), we propose in the $4\times 4$ texture the following mixing matrix:

%Eq. 2
\begin{equation} 
U\! = \! \left( \!\begin{array}{cccc} \, & \, & \, & 0 \\ \, & U^{(3)} & \, & 0 \\  \, & \, & \, &  0  \\ 0  & 0  & 0  & 1 \end{array}\! \right) \left( \!\begin{array}{cccc} \;\,c_{14}  & 0  & 0 & s_{14} \\ \;\,0 & 1 & 0 & 0 \\ \;\,0 & 0 & 1 & 0 \\ -s_{14} & 0 & 0 & c_{14} \end{array} \!\right)\! = \!\left( \!\begin{array}{cccc} \;\,c_{14}/\sqrt{2} & \,1/\sqrt{2} & 0 & \;\,s_{14}/\sqrt{2} \\ -\!c_{14}/2\, & \;\,1/2 & 1/\sqrt{2} & -s_{14}/2\, \\ \;\,c_{14}/2 & \!\!-1/2 & 1/\sqrt{2}  & \;\,s_{14}/2 \\ -s_{14} & \;\,0 & 0 & \;\,c_{14} \end{array}\! \right)
\end{equation} 

\vspace{-0.2cm}

\ni with $ c_{14} = \cos \theta_{14}$ and $ s_{14} = \sin \theta_{14}$ (note that in Eq. (2) only $s_{12}$, $s_{23}$ and $s_{14}$ of all $s_{ij}$ with $i,j = 1,2,3,4 \,,\;i<j$ are nonzero).The unitary transformation describing the mixing of four neutrinos $\nu_\alpha = \nu_e\,, \,\nu_\mu\,, \,\nu_\tau \,,\, \nu_s $ is inverse to the form 

\vspace{-0.2cm}

%Eq. 3
\begin{equation}
\nu_\alpha = \sum_i U_{\alpha i} \nu_i \;,
\end{equation}

\ni where $\nu_i = \nu_1\,, \,\nu_2\,, \,\nu_3\,,\, \nu_4 $ denote four massive neutrino states carrying the masses $m_i = m_1\,, \,m_2\,, \,m_3\,,\, m_4 $. Here, $U = \left(U_{\alpha i } \right)\,,\,\alpha = e\,,\,\mu\,,\,\tau\,,\,s$ and $i = 1,2,3,4$. Of course, $U^\dagger = U^{-1}$ and also $U^* = U$, so that a tiny CP violation is ignored.

In the representation, where the mass matrix of three charged leptons $e^- \,,\, \mu^- \,,\, \tau^-  $ is diagonal, the $4 \times 4$ neutrino mixing matrix $ U $ is at the same time the diagonalizing matrix for the $4 \times 4$ neutrino mass matrix $M = \left(M_{\alpha \beta } \right) $:

\vspace{-0.2cm}

%Eq. 4
\begin{equation} 
U^\dagger M U = {\rm diag}(m_1\,,\,m_2\,,\,m_3\,,\,m_4)\;,
\end{equation} 

\ni where by definition $ m^2_1 \leq m^2_2 \leq m^2_3$ and {\it either} $ m^2_4 \leq m^2_1$ {\it or} $ m^2_3 \leq m^2_4$. Then, due to the formula $ M_{\alpha \beta} = \sum_i U_{\alpha i} m_i U^*_{\beta i}$ we obtain

\vspace{-0.2cm}

%Eq. 5
\begin{eqnarray} 
M_{e e} & = & \frac{1}{2}\left( c^2_{14} m_1 + s^2_{14} m_4 + m_2\right)  \;, \nonumber \\
M_{e \mu} & = & - M_{e \tau} = - \frac{1}{2\sqrt{2}} \left( c^2_{14} m_1 + s^2_{14} m_4 - m_2 \right)\;, \nonumber \\ 
M_{\mu \mu} & = & M_{\tau \tau} = \frac{1}{2} \left[\frac{1}{2} \left( c^2_{14} m_1 + s^2_{14} m_4 + m_2\right) + m_3 \right] = M_{e e} + M_{\mu \tau}\;, \nonumber \\ 
M_{\mu \tau} & = & - \frac{1}{2} \left[\frac{1}{2} \left( c^2_{14} m_1 + s^2_{14} m_4 + m_2\right) - m_3 \right] \;, \nonumber \\ 
M_{e s} & = & - \frac{1}{\sqrt{2}}\, c_{14}\, s_{14} \left( m_1- m_4 \right) \;, \nonumber \\ 
M_{\mu s} & = & - M_{\tau s} = \frac{1}{2}\,c_{14}\, s_{14}\left(m_1 - m_4 \right) = - \frac{1}{\sqrt{2}} M_{e s}\;, \nonumber \\  
M_{s s} & = & s^2_{14} m_1 + c^2_{14} m_4 \;.
\end{eqnarray}

\vspace{-0.2cm}

\ni Of course, $M^\dagger = M$ and also $M^* = M$. From Eqs. (5) we find that 

\vspace{-0.2cm}

%Eq. 6
\begin{eqnarray} 
m_{1,4}\;\,{\rm or}\;\,m_{4,1} & = & \frac{M_{e e} - M_{e \mu}\sqrt{2} + M_{s s}}{2} \pm \sqrt{ \left( \frac{M_{e e} - M_{e \mu}\sqrt{2} - M_{s s}}{2} \right)^{2} + 2 M^2_{e s} }\,, \nonumber \\ m_{2} & = & M_{e e} + M_{e \mu}\sqrt{2}\;\;\;\;,\;\;\;\; m_3\;\; = \;\;M_{\mu \mu} + M_{ \mu \tau}
\end{eqnarray}

\ni if $m_4 \leq m_1$ or $m_1 \leq m_4$, respectively, and 

\vspace{-0.2cm}

%Eq. 7
\begin{equation} 
(2c_{14}s_{14})^2 = \frac{8M^2_{e s}}{(M_{e e} - M_{e \mu}\sqrt{2} - M_{s s})^2 + 8 M^2_{e s}}\; .
\end{equation}

\ni Obviously, $ m_1 + m_2 + m_3 + m_4 = M_{e e} + M_{ \mu \mu} +  M_{\tau \tau} + M_{s s}$ as $  M_{e e} = M_{\mu \mu} - M_{\mu \tau}$ and $ M_{\mu \mu} = M_{\tau \tau}$.

Due to the mixing of four neutrino fields described in Eq. (3), neutrino states mix according to the form

%rownanie 8
\begin{equation} 
|\nu_\alpha \rangle = \sum_i U^*_{\alpha i} |\nu_i \rangle\,.
\end{equation}

\ni This implies the following familiar formulae for probabilities of neutrino oscillations $ \nu_\alpha \rightarrow \nu_\beta $ on the energy shell:

%Eq. 9
\begin{equation} 
P(\nu_\alpha \rightarrow \nu_\beta) = |\langle \nu_\beta| e^{i PL} |\nu_\alpha  \rangle |^2 = \delta _{\beta \alpha} - 4\sum_{j>i} U^*_{\beta j} U_{\beta i} U_{\alpha j} U^*_{\alpha i} \sin^2 x_{ji} \;,
\end{equation}

\ni valid if the quartic product $ U^*_{\beta j} U_{\beta i} U_{\alpha j} U^*_{\alpha i} $ is real, what is certainly true when a possible CP violation can be ignored (then $U^*_{\alpha i} = U_{\alpha i} $). Here,

%Eq. 10
\begin{equation} 
x_{ji} = 1.27 \frac{\Delta m^2_{ji} L}{E} \;\;,\;\;  \Delta m^2_{ji} = m^2_j - m^2_i
\end{equation}

\ni with $\Delta m^2_{ji}$, $L$ and $E$ measured in eV$^2$, km and GeV, respectively ($L$ and $E$ denote the experimental baseline and neutrino energy, while $ p_i = \sqrt{E^2 - m_i^2} \simeq E -  m^2_i/2E $ are eigen\-states of the neutrino momentum $P$).

With the use of oscillation formulae (9), the proposal (2) for the $ 4\times 4$ neutrino mixing matrix leads to the probabilities

%Eq. 11
\begin{eqnarray}
P(\nu_e \rightarrow \nu_e) & \simeq & 1 -  c^2_{14} \sin^2 x_{21} -  \left(1+ c^2_{14}\right) s^2_{14} \sin^2 x _{41} \;, \nonumber \\
P( \nu_\mu \rightarrow \nu_\mu) & = & P( \nu_\tau \rightarrow \nu_\tau)  \simeq 1 - \frac{1}{4} c^2_{14} \sin^2 x _{21} - \frac{1}{2}(1 + c^2_{14}) \left( \sin^2 x _{32} +\frac{1}{2} s_{14}^2 \sin^2 x _{41}\right)  \nonumber \\ & & - \frac{1}{2} s_{14}^2 \sin^2 x _{43} \;, \nonumber \\ 
P( \nu_\mu \rightarrow \nu_e) & = & P( \nu_\tau \rightarrow \nu_e) \simeq \frac{1}{2} \left( c^2_{14} \sin^2 x_{21} + s^4_{14}  \sin^2 x _{41}\right) \;, \nonumber \\
P( \nu_\mu \rightarrow \nu_\tau) & \simeq & - \frac{1}{4} c^2_{14} \sin^2 x_{21} + \frac{1}{2} (1+ c^2_{14}) \left( \sin^2 x _{32} - \frac{1}{2} s_{14}^2 \sin^2 
x _{41} \right) \nonumber \\ & &  + \frac{1}{2} s_{14}^2 \sin^2 x _{43} 
\end{eqnarray}

\ni in the approximation, where $ m^2_1 \simeq m^2_2 $ (and both are different enough from $ m^2_3 $ and $m^2_4 $). The probabilities involving the sterile neutrino $\nu_s $ read: 

%Eq. 12
\begin{eqnarray}
P(\nu_\mu \rightarrow \nu_s) & = & P(\nu_\tau \rightarrow \nu_s) = (c_{14} s_{14})^2 \sin^2 x _{41} \;, \nonumber \\
P(\nu_e \rightarrow \nu_s) & = & 2 (c_{14} s_{14})^2 \sin^2x _{41} \;, \nonumber \\ 
P(\nu_s \rightarrow \nu_s) & = & 1 - (2 c_{14} s_{14})^2 \sin^2 \!x _{41}\;.
\end{eqnarray}

If $\Delta m^2_{21} \ll |\Delta m^2_{41}|$ ({\it i.e.}, $x_{21} \ll |x_{41}|$) and [1,6,7]

%Eq. 13
\begin{equation} 
\Delta m^2_{21} = \Delta m^2_{\rm sol} \sim (10^{-5}\;{\rm or}\;10^{-7}\;{\rm or}\; 10^{-10})  \;{\rm eV}^2  \;,
\end{equation} 

\ni then, under the conditions of solar experiments the first Eq. (11) gives 

%Eq. 14
\begin{equation} 
P(\nu_e \rightarrow \nu_e)_{\rm sol} \simeq 1 -  c^2_{14} \sin^2 (x _{21})_{\rm sol} - \frac{1}{2} (1+c_{14}^2) s_{14}^2  
\end{equation} 

\ni with the estimate

%Eq. 15
\begin{equation} 
c^2_{14} = \sin^2 2\theta_{\rm sol} \sim (0.66\;{\rm or}\;0.97\;{\rm or}\;0.80) \,,\,\frac{1}{2} (1 + c^2_{14}) s^2_{14} \sim (0.28\;{\rm or}\;0.030\;{\rm or}\;0.18)\,,
\end{equation} 

\ni if the constant term in Eq. (14) can be considered as a small perturbation of the usual two-flavor formula. In Eqs. (13) and (15) the recent estimation [7] for the LMA or LOW or VAC solar solution, respectively, is used. Note that

%Eq. 16
\begin{equation} 
P(\nu_e \rightarrow \nu_e)_{\rm sol} \simeq 1 - P(\nu_e \rightarrow \nu_\mu)_{\rm sol} - P(\nu_e \rightarrow \nu_\tau)_{\rm sol} - (c_{14} s_{14})^2
\end{equation} 

\ni with $(c_{14} s_{14})^2 \sim (0.22\;{\rm or}\;0.029\;{\rm or}\;0.16)$.

If $\Delta m^2_{21} \ll \Delta m^2_{32} \ll |\Delta m^2_{41}|\,,\, |\Delta m^2_{43}|$ ({\it i.e.}, $x_{21} \ll x_{32} \ll |x_{41}| \,,\, |x_{43}|$) and [1]

%Eq. 17
\begin{equation} 
\Delta m^2_{32} = \Delta m^2_{\rm atm} \sim 3 \times10^{-3}\;{\rm eV}^2 \;, 
\end{equation} 

\ni then for atmospheric experiments the second Eq. (11) leads to

%Eq. 18
\begin{equation} 
P(\nu_\mu \rightarrow \nu_\mu)_{\rm atm} \simeq  1 - \frac{1}{2}(1 + c^2_{14}) \sin^2 (x _{32})_{\rm atm} - \frac{1}{8} (3+c^2_{14}) s_{14}^2
\end{equation} 

\ni with the prediction

%Eq. 19
\begin{equation} 
\sin^2 2\theta_{\rm atm} = \frac{1}{2}(1 + c^2_{14}) \sim (0.83\;{\rm or}\; 0.99\; {\rm or}\;0.90)\;,\; \frac{1}{8} (3+c^2_{14}) s_{14}^2 \sim (0.16\;{\rm or}\;0.015 \;{\rm or}\;0.095)
\end{equation} 

\ni following from the value (15) of $c^2_{14}$, if again the constant term in Eq. (18) can be considered as a small perturbation. Notice that

%Eq. 20
\begin{equation} 
P(\nu_\mu \rightarrow \nu_\mu)_{\rm atm} \simeq 1 - P( \nu_\mu \rightarrow \nu_\tau)_{\rm atm} - \frac{1}{4}(1 + c_{14}^2) s_{14}^2 
\end{equation} 

\ni with $(1 + c^2_{14}) s_{14}^2/4 \sim (0.14\;{\rm or}\;0.015 \;{\rm or}\;0.09)$.

Eventually, if $\Delta m^2_{21} \ll |\Delta m^2_{41}|$ ({\it i.e.}, $x_{21} \ll  |x_{41}| $) and [2]

%Eq. 21
\begin{equation} 
|\Delta m^2_{41}| = \Delta m^2_{\rm LSND} > 0.1\,,\,{\it e.g.} \sim 1  \;{\rm eV}^2 \;,
\end{equation} 

\ni then for the LSND accelerator experiment the third Eq. (11) implies

%Eq. 22
\begin{equation} 
P( \nu_\mu \rightarrow \nu_e)_{\rm LSND}  \simeq \frac{1}{2}s^4_{14} \sin^2 (x _{41})_{\rm LSND}
\end{equation} 

\ni with the prediction

%Eq. 23
\begin{equation} 
\sin^2 2\theta_{\rm LSND} = \frac{1}{2} s^4_{14} \sim (0.058\;{\rm or}\;4.5 \times 10^{-4} \;{\rm or}\;0.020) 
\end{equation} 

\ni inferred from the value (15) of $ c^2_{14}$. Note that

%Eq. 24
\begin{equation} 
P(\nu_\mu\, \rightarrow\, \nu_e)_{\rm LSND} \simeq \frac{1}{2} \left( \frac{s_{14}}{c_{14}} \right)^2  P( \nu_\mu \rightarrow \nu_s)_{\rm LSND}
\end{equation} 

\ni with $\frac{1}{2}(s_{14}/c_{14})^2 \sim (0.13\;{\rm or}\;4.8 \times 10^{-4} \;{\rm or}\;0.031)$.

It may be worthwhile to remark that if there were $ c^2_{14} =  \sin^2 2\theta_{\rm sol} \sim 0.9 $ (as in the case of older estimations for the LOW solar solution [1]), our predictions would be $ \sin^2 2\theta_{\rm atm}= \frac{1}{2}(1 + c^2_{14}) \sim 0.95$ and $\sin^2 2\theta_{\rm LSND} = \frac{1}{2} s^4_{14} \sim 5 \times 10^{-3}$.

Concluding, we can say that Eqs. (14), (18) and (22) are not inconsistent with solar, atmospheric and LSND experiments. All three depend on one common correlating parameter $c^2_{14}$, implying $c^2_{14} = \sin^2 2\theta_{\rm sol} \sim (0.66\,{\rm or}\,0.97 \,{\rm or}\,0.80)$, $\sin^2 2\theta_{\rm atm} = \frac{1}{2}(1 + c^2_{14}) \sim (0.83\,{\rm or}\,0.99 \,{\rm or}\,0.90)$ and $\sin^2 2\theta_{\rm LSND} = \frac{1}{2}s^4_{14} \sim (0.058\,{\rm or}\,4.5\times 10^{-4} \,{\rm or}\,0.020)$. They depend also on three different mass-square scales $\Delta m^2_{21} = \Delta m^2_{\rm sol} \sim (10^{-5}\,{\rm or}\,10^{-7} \,{\rm or}\, 10^{-10}) \,{\rm eV}^2$, $\Delta m^2_{32} = \Delta m^2_{\rm atm} \sim 3\times 10^{-3} \,{\rm eV}^2 $ and $|\Delta m^2_{41}| = \Delta m^2_{\rm LSND} \sim 1 \,{\rm eV}^2$. Here, the LMA or LOW or VAC solar solution [7] is accepted (perturbatively). Note that in Eqs. (14) and (18) there are constant terms which modify moderately the usual two--flavor formulae. Any LSND--type accelerator project, in contrast to the solar and atmospheric experiments, investigates a small $\nu_\mu \rightarrow \nu_e$ oscillation effect caused possibly by the sterile neutrino $\nu_s$. Thus, this effect (if it exists) plays the role of a small "\,$\!$sterile" ~perturbation of the basic bimaximal texture for three active neutrinos $ \nu_e\,, \,\nu_\mu\,, \,\nu_\tau $. Of course, if $s_{14}$ were zero, such an LSND effect would not exist and both solar $\nu_e \rightarrow \nu_e$ and atmospheric $\nu_\mu \rightarrow \nu_\mu $ oscillations would be maximal. So, from the standpoint of our texture (2), the estimated not full maximality of solar $\nu_e \rightarrow \nu_e$ oscillations may be considered as an argument for the existence of the LSND effect.

The final results (14), (18) and (22) follow from the first three oscillation formulae (11), if {\it either}

%Eq. 25
\begin{equation} 
 m^2_4 \ll m^2_1 \simeq m^2_2 \simeq m^2_3 
\end{equation} 

\ni with

%Eq. 26
\begin{equation} 
m^2_1 \sim 1\;\;{\rm eV}^2 \,,\,  m^2_4 \ll 1\;\;{\rm eV}^2 \,,\, \Delta m^2_{21} \sim (10^{-5}\;{\rm or}\;10^{-7} \;{\rm or}\; 10^{-10})\;{\rm eV}^2 \,,\, \Delta m^2_{32} \sim 3 \times 10^{-3} \;{\rm eV}^2 
\end{equation} 

\ni {\it or}

%Eq. 27
\begin{equation} 
m^2_1 \simeq m^2_2 \ll m^2_3 \ll m^2_4 
\end{equation} 

\ni with

%Eq. 28
\begin{equation} 
m^2_1 \ll 1\;\;{\rm eV}^2\,,\, m^2_4 \sim 1\;{\rm eV}^2 \,,\, \Delta m^2_{21} \sim (10^{-5}\;{\rm or}\;10^{-7} \;{\rm or}\; 10^{-10}) \;{\rm eV}^2 \,,\, \Delta m^2_{32} \sim 3\times 10^{-3} \;{\rm eV}^2 \,.
\end{equation} 

\ni In both cases $\Delta m^2_{21} \ll \Delta m^2_{32} \ll |\Delta m^2_{41}| \sim 1\;{\rm eV}^2 $. The first case of $ m^2_4 \ll m^2_1 \sim 1 \;{\rm eV}^2 $, where the neutrino mass state $ \nu_4$ induced by the sterile neutrino $\nu_s$ gets a vanishing mass, seems to be more natural than the second case of $ m_3^2 \ll m^2_4 \sim 1 \;{\rm eV}^2 $, where such a state gains a considerable amount of mass $\sim 1 \;{\rm eV} $ "for nothing". This is so, unless one believes in the liberal maxim "whatever is not forbidden is allowed". Note that in the first case the neutrino mass states $ \nu_1\,, \,\nu_2\,, \,\nu_3 $ get their considerable masses $\sim 1 \;{\rm eV}$ through spontaneously breaking the electroweak SU(2)$_L \times$U(1) symmetry which, if it were not broken, would forbid these masses.

Finally, for the Chooz reactor experiment [9], where it happens that $(x_{ji})_{\rm Chooz} \simeq (x_{ji})_{\rm atm}$, the first Eq. (11) predicts 

%Eq. 29
\begin{equation} 
P(\bar{\nu}_e \rightarrow\, \bar{\nu}_e)_{\rm Chooz}  \simeq  P( \bar{\nu}_e \rightarrow \bar{\nu}_e)_{\rm atm} \simeq 1 -  \frac{1}{2} (1 + c^2_{14}) s^2_{14}   
\end{equation} 

\ni with $\frac{1}{2}(1 + c^2_{14})s^2_{14} \sim (0.28\;{\rm or}\;0.030 \;{\rm or}\; 0.18) $ from Eq. (15). Here,  $\sin^2 (x_{41})_{\rm Chooz} = \frac{1}{2}$ since $ |(x_{41})_{\rm Chooz}| \simeq |(x_{41})_{\rm atm}| \gg (x_{32})_{\rm atm} \sim 1$ with $|\Delta m^2_{41}| \gg \Delta m^2_{32}$. 

In terms of the usual two--flavor formula, the negative result of Chooz experiment excludes the disappearance process of reactor $ \bar{\nu}_e$'s for moving $(1\! + \!c^2_{14}) s^2_{14} \equiv \sin^2 2\theta_{\rm Chooz} \stackrel{>}{\sim}\! 0.1$, when the range of moving $|\Delta m^2_{41}| \equiv \Delta m^2_{\rm Chooz} \stackrel{>}{\sim} 0.1\,{\rm eV}^2 $ is considered ( then $\Delta m^2_{\rm Chooz} \gg \Delta m^2_{\rm atm} \sim 3\times 10^{-3} \,{\rm eV}^2$, implying $\sin^2 x_{\rm Chooz} = \frac{1}{2}$). Thus, the non\-observation of Chooz effect for reactor $\bar{\nu}_e$'s in the above parameter ranges leads to $ (1 + c^2_{14})s^2_{14} \stackrel{<}{\sim} 0.1$ and hence, to the upper bound $\sin^2 2\theta_{\rm LSND} \equiv \frac{1}{2} s^4_{14} \stackrel{<}{\sim} 1.3\times  10^{-3}$, {\it when} $\Delta m^2_{\rm LSND}\equiv |\Delta m^2_{41}| \equiv \Delta m^2_{\rm Chooz} \stackrel{>}{\sim} 0.1\;{\rm eV}^2 $. It means that $\sin^2 2\theta_{\rm LSND}$, constrained by Chooz (in our four-neutrino texture), lies outside the parameter region suggested at 90\% CL by the {\it existing} LSND data [2], {\it if} the KARMEN2 results [2] excluding a large part of this region are taken into account (in fact, in this corrected region $\sin^2 2\theta_{\rm LSND} \stackrel{>}{\sim} 2\times 10^{-3} $). But, at 99\% CL, this may be not true, allowing for $\Delta m^2_{\rm LSND} \stackrel{>}{\sim}1\;{\rm eV}^2 $ (as, then, in the {\it existing} LSND parameter region $\sin^2 2\theta_{\rm LSND} \stackrel{>}{\sim} 8\times 10^{-4} $). At any rate, among three solar neutrino solutions considered here [7], only the LOW solution is consistent with the Chooz bound [{\it cf.} Eq. (23)]. Also the value $\sin^2 2\theta_{\rm LSND} \sim 5 \times 10^{-3}$, mentioned before as corresponding to older estimations for LOW solar solution [1]: $\sin^2 2\theta_{\rm sol} \sim 0.9$, is eliminated by this bound. 

However, the Chooz-allowed, LOW-induced value  $\sin^2 2\theta_{\rm LSND} \sim 4.5 \times 10^{-4}$ corresponding to $\sin^2 2\theta_{\rm sol} \sim 0.97$ is situated (in contrast to $\sin^2 x_{\rm LSND} \sim 5 \times 10^{-3}$) outside the parameter region implied by the {\it existing} LSND data [2] (even at 99\% CL). 

Of course, the existence of Chooz bound for the LSND effect and of the relation of the latter to  solar neutrino solutions is caused by the {\it correlations} between different neutrino-oscillation modes connected through the parameter $s^2_{14}$ appearing in our four-neutrino texture [{\it cf.} Eqs. (11)]. In fact, the identities $ \frac{1}{2}s^4_{14} = \frac{1}{2} (1 - c^2_{14})^2\, ,\, (1 + c^2_{14}) s^2_{14} = 1 - c^4_{14}$ and $ c^2_{14} + \frac{1}{2} (1+c^2_{14})s^2_{14} + \frac{1}{2}s^4_{14} = 1$ can be translated into the correlations 

%Eq. 30
\begin{equation}
\sin^2 2 \theta_{\rm LSND}   = \frac{1}{2}(1 - \sin^2 2 \theta_{\rm sol})^2 \;\;,\;\;\sin^2 2 \theta_{\rm Chooz} = 1 - \sin^4 2 \theta_{\rm sol}  
\end{equation}

\ni and the sum rule

%Eq. 31
\begin{equation}
\sin^2 2 \theta_{\rm sol} + \frac{1}{2}\sin^2 2 \theta_{\rm Chooz} + \sin^2 2 \theta_{\rm LSND}  = 1  
\end{equation}

\ni for three neutrino-oscillation amplitudes (each in the reasonable two-flavor approximation). The sum rule (31) can be derived also from the probability summation relation $\sum_\beta P(\nu_e \rightarrow \nu_\beta) = 1$ (with $ \beta = e\,,\,\mu\,,\,\tau\,,\,s$) considered under the assumption $m^2_1 \simeq m^2_2 $ for solar $\nu_e$'s (when $|(x_{41})_{\rm sol}| \gg (x_{21})_{\rm sol} \simeq \pi/2$).

We can see that, {\it when} accepting the present Chooz results, we stand with our four-neutrino texture before the  alternative: {\it either} there is no LSND effect at all (then $\sin^2 2\theta_{\rm LSND} \equiv \frac{1}{2} s^4_{14} = 0$ and we are left with the three-neutrino bimaximal texture [4,8]), {\it or} this effect exists all right, but at a point in the parameter space, where the oscillation amplitude $\sin^2 2\theta_{\rm LSND}$ is shifted ({\it versus} the existing LSND data) towards a smaller value $\sin^2 2\theta_{\rm LSND} \equiv \frac{1}{2} s^4_{14} \stackrel{<}{\sim} 1.3\times  10^{-3}$ (though >0). Note that, if $\sin^2 2\theta_{\rm LSND} \equiv \frac{1}{2} s^4_{14}$ was at the Chooz bound value $1.3\times  10^{-3}$, then $\sin^2 2\theta_{\rm sol} \equiv c^2_{14}$ would be at 0.95. If, rather, $\sin^2 2\theta_{\rm LSND} \equiv \frac{1}{2} s^4_{14}$ was at the value $8\times 10^{-4}$ equal to its {\it existing} LSND lower limit at 99\% CL, then $\sin^2 2\theta_{\rm sol} \equiv c^2_{14}$ would be at 0.96. Of course, we should keep in mind the fact that the present estimates for $\sin^2 2\theta_{\rm sol}$ (and even more for $\sin^2 2\theta_{\rm LSND}$) are preliminary and, in fact, very fragile.

From the neutrinoless double $\beta$ decay, not observed so far,  the experimental bound $ \overline{M}_{ee} \equiv |\sum_i U^2_{e i} m_i| \stackrel{<}{\sim} [0.4\, (0.2)-1.0\, (0.6)]$ eV follows [10] (here, $U^2_{ei}$ appears even if $U^*_{ei} \neq U_{ei}$). On the other hand,for {\it e.g.} the LOW solar solution the first Eq. (5) gives

%Eq. 32
\begin{equation} 
\overline{M}_{ee} = |{M}_{ee} |\sim \frac{1}{2}|0.97 m_1 + 0.03 m_4 + m_2|\;,
\end{equation}

\ni what in the case of Eq. (25) with $m_1 \!\sim\! \pm1$ eV and $ m_2 \!\sim\! 1$ eV or Eq. (27) with $ |m_4| \!\sim\! 1$~eV  leads to the estimation $\overline{M}_{ee} \sim (0.99,\, 0.015)$ eV or $\overline{M}_{ee} \sim 0,015$ eV, respectively (putting $\overline{M}_{ee} = |M_{ee}|$ in Eq. (32) we ignore a possible violation of CP: we get $U^*_{ei} = U_{ei}$, since $ M_{ee} = \sum_i | U_{e i}|^2 m_i $). 

In the case of neutrinoless double $\beta$ decay we assume, of course, that in our texture four flavor neutrinos $\nu_\alpha = \nu_e\,, \,\nu_\mu\,, \, \nu_\tau \,, \,\nu_s $ are Majorana fermions $\nu_\alpha  = \nu_{\alpha L} + \left( \nu_{\alpha L} \right)^c $, where $ \nu_{\alpha R} = (\nu_{\alpha L})^c = (\nu^c_{\alpha})_ R$ and $\nu_{\alpha} =  \nu^c_{\alpha}$. Then, the neutrino mass term in the Lagrangian density,

%rownanie 33
\begin{equation}
-{\cal L}^{\rm (light)}_{\rm mass} \equiv \frac{1}{2}\sum_{\alpha \beta} \bar{\nu}_\alpha\, M_{\alpha \beta}\, \nu_\beta
\end{equation} 

\ni with $M = (M_{\alpha \beta})$ given in Eq. (5), is a Majorana lefthanded mass term, as $\nu_\alpha$ are built up of $\nu_{\alpha L}$ [and $ (\nu_{\alpha L})^c$].

It may happen, however, that in this case, beside four flavor neutrinos $\nu_{\alpha}$ corresponding to four {\it light} mass neutrino states  $ \nu_i = \nu_1 \,,\,\nu_2 ,\,\nu_3 \,,\, \nu_4$, there exist four other flavour neutrinos $ \nu'_\alpha = \nu'_e\,, \,\nu'_\mu\,, \,\nu'_\tau \,,\, \nu'_s$, being also Majorana fermions $\nu'_\alpha  = \nu'_{\alpha R} + \left( \nu'_{\alpha R} \right)^c $, where $ \nu'_{\alpha L} = (\nu'_{\alpha R})^c = (\nu'^c_{\alpha})_ L$ and $\nu'_{\alpha} =  \nu'^c_{\alpha}$, but connected this time with four {\it heavy} mass neutrino states $ \nu'_i = \nu'_1 \,,\,\nu'_2 ,\,\nu'_3 \,,\, \nu'_4 $. Thus, the latter may be practically decoupled from the former light mass neutrino states. In other words, a {\it four-neutrino} seesaw mechanism may work, "\,$\!$explaining"\, the lightness of $\nu_i$ {\it versus} $\nu'_i$:

%Eq. 34
\begin{eqnarray}
M_{\alpha \beta} \equiv M^{\rm (light)}_{\alpha \beta} \simeq M^{(L)}_{\alpha \beta}  - \left( M^{(D)} M^{(R)\,-1} M^{(D)} \right)_{\alpha \beta}\!\!\!\! & \simeq & \!\!\!\! - \left( M^{(D)} M^{(R)\,-1} M^{(D)} \right)_{\alpha \beta} \;, \nonumber \\  M'_{\alpha \beta} \equiv M^{\rm (heavy) }_{\alpha \beta} \simeq M^{(R)}_{\alpha \beta}\!\!\!\! &,& 
\end{eqnarray}

\ni where $M^{(L)} \ll M^{(D)} \ll M^{(R)}$ in the perturbative sense. These $4\times 4$ matrices, {\it viz.} Majorana lefthanded, Dirac and Majorana righthanded, respectively, are assumed to be real and symmetric for simplicity. Here, the overall neutrino mass term in the Lagrangian density has the form

% Eq. 35
\begin{equation}
-{\cal L}_{\rm mass} \equiv \frac{1}{2}\sum_{\alpha \beta} (\bar{\nu}_\alpha \,,\, \bar{\nu}'_{\alpha}) \left( \begin{array}{cc} M^{(L)}_{\alpha \beta} & M^{(D)}_{\alpha \beta} \\ M^{(D)}_{\alpha \beta} & M^{(R)}_{\alpha \beta} \end{array} \right) \left( \begin{array}{c} \nu_\beta \\ \nu'_{\beta} \end{array} \right) \nonumber \simeq -{\cal L}^{\rm (light)}_{\rm mass} -{\cal L}^{\rm (heavy)}_{\rm mass} \;.
\end{equation} 

\ni Note that the combinations $\nu^{(D)}_\alpha \equiv \nu_{\alpha L} + \nu'_{\alpha R}$ are formed in the sum of two terms with $M_{\alpha \beta}^{(D)} $ in Eq. (35) and play there the role of four Dirac flavor neutrinos.

For three conventional $\alpha = e\,,\,\mu\,,\,\tau $ the Majorana neutrinos $\nu_\alpha = \nu_{\alpha L} + (\nu_{\alpha L})^c = \nu_{\alpha L}^{(D)} + (\nu_{\alpha L}^{(D)})^c$ and $ \nu'_\alpha = \nu'_{\alpha R} + (\nu'_{\alpha R})^c = \nu_{\alpha R}^{(D)} + (\nu_{\alpha R}^{(D)})^c $ were denoted also by $\nu^{(a)}_\alpha$ and $\nu^{(s)}_\alpha$, and called active and conventional-sterile neutrinos, while the Dirac neutrinos $\nu^{(D)}_\alpha$ were written simply as $\nu_\alpha$ ({\it cf. e.g.} Ref. [5]). Then, $\nu'_{\alpha R} =\nu^{(D)}_{\alpha R}$ were written as $\nu_{\alpha R}$.

The sterile neutrino $\nu_s$ --- the extra flavor neutrino considered in this paper beside the active neutrinos $ \nu_e\,, \,\nu_\mu\,, \,\nu_\tau $  --- differs in an obvious way from any of four possible conventional-sterile neutrinos $ \nu'_e\,, \,\nu'_\mu\,, \,\nu'_\tau $ and $ \nu'_s$ needed for the four-neutrino seesaw mechanism to be realized (and for four Dirac neutrinos $ \nu^{(D)}_e\,, \,\nu^{(D)}_\mu\,, \,\nu^{(D)}_\tau $ and $ \nu^{(D)}_s$ to be defined).

\vfill\eject 

~~~
\vspace{0.3cm}

\centerline {\bf Appendix: Allowing for nonzero $ s_{13}$}

\vspace{0.3cm}

If we maintain two maximal mixings $c_{12} = 1/\sqrt{2} = s_{12}$ and $c_{23} = 1/\sqrt{2} = s_{23}$, but decide to allow $ s_{13}$ to be nonzero (though small, as may be expected), then Eqs. (1) and (2) take the forms

% (A 1)
$$
U^{(3)} =   \left( \begin{array}{ccc} c_{13}/\sqrt{2} & c_{13}/\sqrt{2} & s_{13} \\
- (1+s_{13})/{2}\, & \;\,(1-s_{13})/2 & c_{13}/\sqrt{2} \\ \;\,(1-s_{13})/2 & -(1+s_{13})/2\, & c_{13}/\sqrt{2}\end{array} \right)  \eqno({\rm A\,1})
$$

\ni and

% (A\,2)
$$
U =   \left( \begin{array}{cccc} c_{13}c_{14}/\sqrt{2} & c_{13}/\sqrt{2} & s_{13} & c_{13}s_{14}/\sqrt{2} \\ - (1+s_{13})c_{14}/{2}\, & \;\,(1-s_{13})/2 & c_{13}/\sqrt{2} & -(1+s_{13})s_{14}/2 \\ \;\,(1-s_{13})c_{14}/2 & -(1+s_{13})/2\, & c_{13}/\sqrt{2} & \;\,(1-s_{13}) s_{14}/2 \\ -s_{14} & 0 & 0 & c_{14}
\end{array} \right) \;, \eqno({\rm A\,2})
$$

\ni respectively, where we put the CP violating phase $\,\delta_{13} = 0\,$ [also $\,\delta_{14} = 0\,$, as before in Eq. (2)].

In such a case, the neutrino oscillation formulae (9) lead for $m^2_1 \simeq m^2_2$ to the probabilities

\vspace{-0.2cm}

% (A\,3)
\begin{eqnarray*}
P(\nu_e\;\; \rightarrow \;\;\nu_e) & \simeq & 1 -  c^4_{13}c^2_{14} \sin^2 x_{21} - c^2_{13} \left(1+ c^2_{14}\right)\left( 2s^2_{13}\sin^2 x_{32} + c^2_{13} s^2_{14} \sin^2 x_{41}\right) \\ & &  - \;2 c^2_{13} s^2_{13} s^2_{14} \sin^2 x_{43}\;,  \\
P( \nu_{\mu,\tau} \rightarrow \nu_{\mu,\tau})\!\! & \simeq & 1 - \frac{1}{4}\, c^4_{13}c^2_{14} \sin^2 x _{21} - \left[\frac{1}{2}(1+s^2_{13})(1 + c^2_{14}) \mp s_{13}s^2_{14}\right] \\ & & \times \!\left[ c^2_{13} \sin^2 x _{32}\! +\!\frac{1}{2}(1\!\pm\! s_{13})^2 s_{14}^2 \sin^2 x _{41}\right]\! - \! \frac{1}{2}c^2_{13}(1 \!\pm\! s_{13})^2 s_{14}^2 \sin^2 x _{43} \;,  \\ 
P( \nu_{\mu,\tau} \rightarrow \;\;\nu_e)\! & \simeq & \frac{1}{2}\,c^4_{13}c^2_{14}  \sin^2 x_{21} + c^2_{13}\left[ s^4_{14} \mp s_{13}(1 \!+\! c^2_{14})\right] \\ & & \times \left[\mp s_{13} \sin^2 x _{32} \!+\! \frac{1}{2}(1\!\pm\! s_{13})s^2_{14}\sin^2 x _{41}\right] \pm c^2_{13} s_{13}(1\pm s_{13})s^2_{14} \sin^2 x_{43} \;,  \\ 
P( \nu_\mu\;\,\rightarrow \;\nu_\tau) & \simeq & - \frac{1}{4}\,c^4_{13} c^2_{14} \sin^2 x_{21} + \frac{1}{2}\,c^4_{13} (1+ c^2_{14}) \left( \sin^2 x _{32} - \frac{1}{2} s_{14}^2 \sin^2 
x _{41} \right)  \\ & &  + \frac{1}{2}\,c^4_{13} s_{14}^2 \sin^2 x _{43} \;.
\end{eqnarray*} 

\vspace{-1.52cm}

\begin{flushright}
(A\,3)
\end{flushright}

\ni The probabilities involving the sterile neutrino $\nu_s$ are

%\vspace{-0.2cm}

% (A\,4)
\begin{eqnarray*}
P( \nu_{\mu,\tau} \rightarrow \nu_s) & = & (1\pm s_{13})^2 c^2_{14}s^2_{14} \sin^2 x _{41}\;, \\ 
P(\nu_e\;\, \rightarrow \nu_s)\, & = & 2 c^2_{13} c^2_{14}s^2_{14}\sin^2 x_{41} \;, \\
P( \nu_s\;\, \rightarrow \nu_s)\, & = & 1 - 4 c^2_{14}s^2_{14}  \sin^2 x_{41} \;.  
\end{eqnarray*} 

\vspace{-1.52cm}

\begin{flushright}
(A\,4)
\end{flushright}

\ni The formulae (A 3) and (A 4) are obviously reduced to Eqs. (11) and (12) in the limit of $ s_{13} \rightarrow 0 $. 

In the linear approximation with respect to a small $\,s_{13}\,$, only the probabilities $\; P(\nu_{\mu,\tau}\! \rightarrow \!\nu_{\mu,\tau})$, $P( \nu_{\mu,\tau} \rightarrow \nu_e)$ and $ P(\nu_{\mu,\tau} \rightarrow \nu_s)$ of those given in Eqs. (A~3) and (A~4) get corrections to Eq. (11) and (12):

% (A\,5)
\begin{eqnarray*}
\delta P( \nu_{\mu,\tau} \rightarrow \nu_{\mu,\tau})\! & \simeq & \pm s_{13} s^2_{14} \left(\sin^2 x _{32} - c^2_{14} \sin^2 x _{41} - \sin^2 x _{43}\right) \;, \\ 
\delta P(\nu_{\mu,\tau}  \rightarrow \;\; \nu_e) & \simeq & \mp s_{13} s^2_{14} \left(\sin^2 x _{32} + c^2_{14} \sin^2 x _{41} - \sin^2 x _{43}\right) \;, \\ 
\delta P( \nu_{\mu,\tau}  \rightarrow \;\; \nu_s) & \simeq & \pm 2 s_{13} s^2_{14} c^2_{14} \sin^2 x _{41} \;.  
\end{eqnarray*} 

\vspace{-1.52cm}

\begin{flushright}
(A\,5)
\end{flushright}

\ni Thus, in the linear approximation in $ s_{13}$, when making use of Eqs. (14), (29) and (18), (22) as well as (A 5), we obtain

% (A\,6)
\begin{eqnarray*}
P(\nu_e \rightarrow \nu_e)_{\rm sol}\;\;\, & \simeq & 1 - c^2_{14} \sin^2 (x_{21})_{\rm sol}  - \frac{1}{2}(1+ c^2_{14}) s^2_{14} \;,  \\
P(\bar{\nu}_e \rightarrow \bar{\nu}_e)_{\rm Chooz}\!\! & \simeq & 1 - \frac{1}{2}(1+ c^2_{14}) s^2_{14} \;,  \\
P( \nu_\mu \rightarrow \nu_\mu)_{\rm atm}\, & \simeq & 1 - \frac{1}{2}(1+ c^2_{14}) \sin^2 (x _{32})_{\rm atm} - \frac{1}{8}(3+c^2_{14}) s^2_{14} \\ 
& & +\, s_{13} s^2_{14}\left[ \sin^2 (x _{32})_{\rm atm}   - \frac{1}{2}(1+ c^2_{14})\right] \\ 
& = & 1 - \left[ \frac{1}{2}(1 + c_{14}^2) - s_{13}s^2_{14}\right] \sin^2 (x _{32})_{\rm atm} \\ 
& & - \,\left[ \frac{1}{8}(3+ c_{14}^2) + \frac{1}{2}s_{13}(1 + c^2_{14}) \right] s^2_{14} \;,  \\ 
P( \nu_\mu \rightarrow \nu_e)_{\rm LSND}\!\!\! & \simeq & \frac{1}{2}s^4_{14} \sin^2 (x_{41})_{\rm LSND}  - s_{13} s^2_{14} \left[ c^2_{14}\sin^2 (x_{41})_{\rm LSND} - \sin^2 (x_{43})_{\rm LSND} \right] \\ 
& = &  \left[ \frac{1}{2} s^4_{14} -  s_{13} s^2_{14} c^2_{14}\right] \sin^2 (x_{41})_{\rm LSND} + s_{13}s^2_{14} \sin^2 (x_{43})_{\rm LSND} \;,
\end{eqnarray*} 

\vspace{-1.52cm}

\begin{flushright}
(A\,6)
\end{flushright}

\ni where $\Delta m^2_{21} \ll \Delta m^2_{32} \ll |\Delta m^2_{41}|\,,\, |\Delta m^2_{43}|$ and $ (x_{21})_{\rm sol}\sim 1$, $ (x_{32})_{\rm atm} \sim 1$, $|(x_{41})_{\rm LSND} | \sim 1$. We can see that, in this linear approximation, though the Chooz restriction $(1 + c^2_{14}) s^2_{14} \stackrel{<}{\sim} 0.1$ {\it i.e.}, $ \frac{1}{2} s^4_{14} \stackrel{<}{\sim} 1.3 \times 10^{-3}$, is not changed formally, the resulting amplitude for the possible LSND effect becomes slightly different. In fact, if $|\Delta m^2_{41}| \simeq |\Delta m^2_{43}|$, what holds both in the case of $ m^2_4 \ll m^2_1 \simeq m^2_2 \simeq m^2_3 \sim 1\;{\rm eV}^2$ and $ m^2_1\simeq m^2_2 \ll m^2_3  \ll m^2_4 \sim 1\;{\rm eV}^2$, the last Eq. (A 6) gives 

% (A\,7)
$$
P( \nu_\mu \rightarrow \nu_e)_{\rm LSND} \simeq  \frac{1}{2}s^4_{14} (1+ s_{13}) \sin^2 (x_{41})_{\rm LSND}  \;,
\eqno({\rm A\,7})
$$

\ni leading to the oscillation amplitude

% (A\,8)
$$
\sin^2 2\theta_{\rm LSND} = \frac{1}{2} s^4_{14} (1 + s_{13}) \;,
\eqno({\rm A\,8})
$$

\ni a little larger (smaller) than before with $s_{13} = 0$, if $s_{13} > 0$ ($s_{13} < 0$).

\vfill\eject

~~~~
\vspace{0.5cm}

{\centerline{\bf References}}

\vspace{0.5cm}

{\everypar={\hangindent=0.6truecm}
\parindent=0pt\frenchspacing

{\everypar={\hangindent=0.6truecm}
\parindent=0pt\frenchspacing

~[1]~For a review {\it cf.} E. Kearns, Plenary talk at {\it ICHEP 2000} at Osaka; C.~Gonzales--Garcia, Talk at {\it ICHEP 2000} at Osaka; and references therein.

\vspace{0.2cm}

~[2]~G. Mills, Talk at {\it Neutrino 2000}, {\it Nucl. Phys. Proc. Suppl.} {\bf 91}, 198 (2001);  R.L.~Imlay, Talk at {\it ICHEP 2000} at Osaka; and references therein; for recent results of the KARMEN2 experiment {\it cf.} K.~Eitel, Talk at {\it Neutrino 2000}, {\it Nucl. Phys. Proc. Suppl.} {\bf 91}, 191 (2001).

\vspace{0.2cm}

~[3]~W. Kr\'{o}likowski, {\it Nuovo Cim.} {\bf A 111}, 1257 (1998); {\it Acta Phys. Pol.} {\bf B 31}, 1759 (2000).

\vspace{0.2cm}

~[4]~For a review {\it cf.} F. Feruglio, {\it Acta Phys. Pol.} {\bf B 31}, 1221 (2000); J.~Ellis, Summary of {\it Neutrino 2000}, {\it Nucl. Phys. Proc. Suppl.} {\bf 91}, 503 (2001); and references therein.

\vspace{0.2cm}

~[5]~W. Kr\'{o}likowski, {\it Acta. Phys. Pol.} {\bf B 32}, 1245 (2001); hep--ph/0102016; hep--ph/0103226.

\vspace{0.2cm}

~[6].~M.V.~Garzelli and C.~Giunti, hep--ph/0012247.

\vspace{0.2cm}

~[7]~J.N.~Bahcall, P.I.~Krastev and A.Yu.~Smirnov, hep--ph/0103179.

\vspace{0.2cm}

~[8]~{\it Cf. e.g.},~W. Kr\'{o}likowski, hep--ph/0007255; {\it Acta. Phys. Pol.} {\bf B 32}, 75 (2001).

\vspace{0.2cm}

~[9]~M. Appolonio {\it et al.}, {\it Phys. Lett.} {\bf B 420}, 397 (1998); {\bf B 466}, 415 (1999).

\vspace{0.2cm}

[10].~L.~Baudis {\it et al.}, {\it Phys. Rev. Lett.} {\bf 83}, 41 (1999); {\it cf.} also Review of Particle Physics, {\it Eur. Phys. J.} {\bf C 15}, 1 (2000), p. 363.

\vfill\eject

\end{document}